\documentclass[aps,preprint,groupedaddress,floatfix]{revtex4-1}
\usepackage{graphicx}
\begin{document}

\title{Generalized Vector Dominance Model up to 2 GeV}
\author {
N.N. Achasov$^{\,a}$ \email{achasov@math.nsc.ru}, A.V.
Kiselev$^{\,a,b}$ \email{kiselev@math.nsc.ru}, and A.A. Kozhevnikov$^{\,a,b}$ \email{kozhev@math.nsc.ru}}

\affiliation{ $^a$Laboratory of Theoretical Physics,
 Sobolev Institute for Mathematics, 630090, Novosibirsk, Russia\\
$^b$Novosibirsk State University, 630090, Novosibirsk, Russia}

\date{\today}

\begin{abstract}

The processes $e^+e^-\to \pi^+\pi^-, \omega\pi^0, \eta\pi^+\pi^-, K^+K^-, \pi^+\pi^-\pi^0$ are described simultaneously in the frame of the Generalized Vector Dominance Model (GVDM) allowing model error. The data do not contradict the assumption that the precision of the model we use is not worse than 6\%. This shows that GVDM is an adequate approach to describe processes involving $\rho(770),\omega(787),\phi(1020)$ and their excitations in the energy region below 2 GeV. This work is considered as a first step, next steps should treat more processes and take into account finer effects.

\end{abstract}

\maketitle

\section{Introduction}

Phenomenological Generalized Vector Dominance Model (GVDM) is commonly used to describe experimental data involving vector mesons in the energy region up to 2 GeV and even higher. The existence of the particles taken into account by GVDM, $\rho^{\prime}_1(1450), \rho^{\prime}_2(1700), \omega^{\prime}_1(1420), \omega^{\prime}_2(1650), \phi^{\prime}_1(1680), \phi^{\prime}_2(2170)$ and others, is not questioned \cite{pdg-2020} (notations are same as in Refs. \cite{argus,isoscalars}). Properties of these particles are not known well, but the data coming from VEPP-2000, BESIII and other machines should clarify them.

On the current stage different modifications of GVDM are used to fit each data set on each process separately. Resulted parameters are different and sometimes even contradict each other. But what will happen if one takes some variant of GVDM and tries to describe different processes simultaneously with one common set of parameters? Such an attempt should clarify whether this model is adequate, reveal the precision of the model, and allow to increase the precision by improving the model.

In this paper we present simultaneous description of the currently most precise data on five processes in the energy region below 2 GeV Refs. \cite{babarPP,sndOmPi,sndEtaPP,KK,3Pi}. These processes are $e^+e^-\to \pi^+\pi^-, \omega\pi^0, \eta\pi^+\pi^-$, dominated by isovector vector resonances contribution, and $e^+e^-\to K^+K^-, \pi^+\pi^-\pi^0$, dominated by the isoscalar vectors contribution with sizeable isovector contribution to $e^+e^-\to K^+K^-$ reaction. The model to describe the data is taken from Refs. \cite{argus,isoscalars}) with local improvements.

It turns out that the data could be described very well if one assumes 6\% overall accuracy of GVDM and simultaneously better accuracy in $e^+e^-\to \pi^+\pi^-$ and $e^+e^-\to \omega\pi^0$ in particular. This shows that GVDM can pretend to be rather precise as a common theory. Note that we don't demand the same precision from SU(3) relations, they can be violated more.

By model accuracy we mean a minimal model error, consistent with experimental data. In other words, the hypothesis that all true values are located inside the corridor formed by model error bars should be statistically consistent with data, details are in Secs. \ref{SecChi2} and \ref{dataDescr}.

Of course, there are more processes of interest than we take into account now. Some effects could be taken into account also. We plan to enhance our analysis during next steps.

\section{Description of the reactions}

\subsection{Isovector channels}

The reactions $e^+e^-\to\rho+\rho^{\prime}_1+\rho^{\prime}_2\to \pi^+\pi^-, \omega\pi^0, \eta\pi^+\pi^-$ are described by the expression

\begin{equation}
\sigma(e^+e^-\to\rho+\rho^{\prime}_1+\rho^{\prime}_2\to f)=
\frac{4\pi\alpha^2}{s^{3/2}}\Biggl|\Biggl(\frac{m^2_\rho}{f_\rho},
\frac{m^2_{\rho^{\prime}_1}}{f_{\rho^{\prime}_1}},
\frac{m^2_{\rho^{\prime}_2}}{f_{\rho^{\prime}_2}}\Biggr)G_\rho^{-1}(s)
\pmatrix{g_{\rho f}\cr g_{\rho^{\prime}_1 f}\cr g_{\rho^{\prime}_2 f}
\cr}
\Biggr|^2P_f,
\label{eq1}
\end{equation}
\noindent where $f=\pi^+\pi^-$, $\omega\pi^0$ and $\eta \pi^+\pi^-$; $s$ is the total
center-of-mass energy squared; $\alpha\approx 1/137$. The $\rho\omega$ mixing should be taken into account in the case of the $\pi^+\pi^-$ channel, it will
be done below. For $\rho_i=\rho(770)\mbox{, }
\rho^{\prime}_1\mbox{, }\rho^{\prime}_2$ leptonic coupling constants $f_{\rho_i}$ are connected to the widths as usual:
\begin{equation}
\Gamma_{\rho_i\to e^+e^-}=\frac{4\pi\alpha^2}{3f_{\rho_i}^2}m_{\rho_i}.
\label{eq1a}
\end{equation}
The factor $P_f$ for the mentioned final states reads, respectively
\begin{equation}
P_f\equiv P_f(s)=
\frac{2}{3s}q^3_{\pi\pi}\mbox{, }\frac{1}{3}q^3_{\omega\pi}\mbox{, }
\frac{1}{3}\langle q^3_{\rho\eta}\rangle\cdot\frac{2}{3},
\label{eq2}
\end{equation}
where
\begin{equation}
q_{ij}\equiv q(M,m_i,m_j)=\frac{1}{2M}\sqrt{[M^2-(m_i-m_j)^2]
[M^2-(m_i+m_j)^2]}
\label{eq6}
\end{equation}
is the momentum of either particle $i$ or $j$, in the rest frame of the decaying particle.

In case of $\eta\pi^+\pi^-$ the $\langle q^3_{\rho\eta}\rangle$ is
\begin{equation}
\langle q^3_{\rho\eta}\rangle=\int\limits_{(2m_\pi)^2}
^{(\sqrt{s}-m_\eta)^2}dm^2
\rho_{\pi\pi}(m)q^3(\sqrt{s},m,m_\eta).
\label{eq2a}
\end{equation}
The function  $\rho_{\pi\pi}(m)$ accounts for the finite
width of the intermediate $\rho(770)$ meson and looks as 
\begin{equation}
\rho_{\pi\pi}(m)=\frac{\frac{1}{\pi}m\Gamma_{\rho\to\pi\pi}(m)}
{(m^2-m^2_\rho)^2+(m\Gamma_{\rho\to\pi\pi}(m))^2},
\label{eq5}
\end{equation}
where $\Gamma_{\rho\to\pi\pi}(m)=\frac{g^2_{\rho\pi\pi}}{6\pi s}q^3_{\pi\pi}$ is the $\pi\pi$ width of the $\rho$ meson. In the approximation Eq. (\ref{eq5}) we take $m_\rho=774$ MeV, $g_{\rho\pi\pi}=5.9$ as in Ref. \cite{argus}. Note that here we have an effective physical $\rho$ meson, and it is not bare non-mixed $\rho$ meson from the matrix below.

The multiplier 2/3 arises in the simplest quark model, namely, the relation between the $\rho\eta$ and $\omega\pi^0$ production amplitudes,
provided the pseudoscalar mixing angle is taken to be $\theta_P=-11^o$. This estimation is not precise, so for $\rho^\prime_{1,2}$ we use the correction factors $y_{\eta i}$, $i=1,2$, such that
\begin{equation}g_{\rho^\prime_i\rho\eta}=y_{\eta i}\sqrt{2/3}g_{\rho^\prime_i\omega\pi}\\,\label{yDef}\end{equation}and require $y_{\eta i}$ to be not far from 1.

The matrix of inverse propagators reads
\begin{equation}
G_\rho(s)=\pmatrix{D_\rho&-\Pi_{\rho\rho^{\prime}_1}&-\Pi_{\rho\rho^{\prime}_2}\cr
-\Pi_{\rho\rho^{\prime}_1}&D_{\rho^{\prime}_1}&-\Pi_{\rho^{\prime}_1
\rho^{\prime}_2}\cr
-\Pi_{\rho\rho^{\prime}_2}&-\Pi_{\rho^{\prime}_1\rho^{\prime}_2}&
D_{\rho^{\prime}_2}\cr}.
\label{eq7}
\end{equation}
It contains the inverse propagators of the unmixed states
\begin{equation}
D_{\rho_i}\equiv D_{\rho_i}(s)=m^2_{\rho_i}-s-i\sqrt{s}\Gamma_{\rho_i}(s),
\label{eq8}
\end{equation}
\noindent 


\noindent where the energy dependent widths $\Gamma_{\rho_i}(s)$ cover the channels $\rho_i\to\pi\pi$, $K\bar K,\omega\pi,$ $K^{\ast}K,\eta\pi^+\pi^-,$ $4\pi$:
\begin{eqnarray}
\Gamma_{\rho_i}(s)&=&\frac{g^2_{\rho_i\pi\pi}}{6\pi s}q^3_{\pi\pi}+\frac{g^2_{\rho_i K^+K^-}}{6\pi s}q^3_{K^+K^-}+\frac{g^2_{\rho_i K^0\bar K^0}}{6\pi s}q^3_{K^0\bar K^0}+
\frac{g^2_{\rho_i\omega\pi}}{12\pi}\biggl(q^3_{\omega\pi}+q^3_{K^{\ast}K}+
\frac{2}{3}\langle q^3_{\rho\eta}\rangle\biggr)+    \nonumber\\
& &{3\over2}
g^2_{\rho_i\rho^0\pi^+\pi^-}W_{\pi^+\pi^-\pi^+\pi^-}(s)+g^2_{\rho_i\rho^+\rho^-}W_{\pi^+\pi^-\pi^0\pi^0}(s)\,,
\label{eq9}
\end{eqnarray}
the channel $K^{\ast}K\pi$ is omitted because it turned out to be consistent with zero in Ref. \cite{argus}; the $4\pi$ phase spaces are

\begin{eqnarray}
W_{\pi^+\pi^-\pi^+\pi^-}(s)&=&\frac{1}{(2\pi)^34s}\int
\limits_{(2m_\pi)^2}^{(\sqrt{s}-2m_\pi)^2}dm^2_1
\rho_{\pi\pi}(m_1)\int\limits_{(2m_\pi)^2}^{(\sqrt{s}-m_1)^2}dm^2_2
\times \nonumber\\
& &\biggl(1+
\frac{q^2(\sqrt{s},m_1,m_2)}{3m^2_1}\biggr)
q(\sqrt{s},m_1,m_2)q(m_2,m_\pi,m_\pi),   \nonumber\\
W_{\pi^+\pi^-\pi^0\pi^0}(s)&=&\frac{1}{2\pi s}\int
\limits_{(2m_\pi)^2}^{(\sqrt{s}-2m_\pi)^2}dm^2_1
\rho_{\pi\pi}(m_1)\int\limits_{(2m_\pi)^2}^{(\sqrt{s}-m_1)^2}dm^2_2
\rho_{\pi\pi}(m_2)q^3(\sqrt{s},m_1,m_1),
\label{phsp}
\end{eqnarray}

\noindent and the nondiagonal polarization operators
$$\Pi_{\rho_i\rho_j}=\mbox{Re}\Pi_{\rho_i\rho_j}+i\mbox{Im}
\Pi_{\rho_i\rho_j}$$
describe the mixing. Their imaginary parts are given by the
unitarity relation and read
\begin{eqnarray}
\mbox{Im}\Pi_{\rho_i\rho_j}&=&\sqrt{s}
\biggl[\frac{g_{\rho_i\pi\pi}g_{\rho_j\pi\pi}}{6\pi s}q^3_{\pi\pi}+
\frac{g_{\rho_i\omega\pi}g_{\rho_j\omega\pi}}{12\pi}
\biggl(q^3_{\omega\pi}+q^3_{K^{\ast}K}+
\frac{2}{3}\langle q^3_{\rho\eta}\rangle\biggr)+     \nonumber\\
& & {3\over2}g_{\rho_i\rho^0\pi^+\pi^-}g_{\rho_j\rho^0\pi^+\pi^-}
W_{\pi^+\pi^-\pi^+\pi^-}(s)+
g_{\rho_i\rho^+\rho^-}g_{\rho_j\rho^+\rho^-}
W_{\pi^+\pi^-\pi^0\pi^0}(s)\biggr].
\label{eq10}
\end{eqnarray}

Real parts of $\Pi_{\rho_i\rho_j}$ are unknown, as in \cite{argus} we assume that they are constant.

The quark model relations were used in Eqs. (\ref{eq9}) and (\ref{eq10}) for the couplings of $\rho_i$ with the vector+pseudoscalar states. The possibility of some violation of these
relations will be included below. The quark model relations for the $\rho(770)\to K\bar K, K^*K$ couplings could be found in Sec. \ref{isosc}.

The $\rho\omega$ mixing is taken into account for the $\pi^+\pi^-$ channel. The term

$$\frac{m^2_\omega\Pi_{\rho\omega}}
{f_\omega D_\rho D_\omega}g_{\rho\pi\pi},$$
is added to the expression in between the modulus sign in Eq. (\ref{eq1}), where
\begin{eqnarray}
D_\omega&\equiv& D_\omega(s)=
m^2_\omega-s-i\sqrt{s}\Gamma_\omega(s),  \nonumber\\
\Gamma_\omega(s)&=&\frac{g^2_{\omega\rho\pi}}{4\pi}W_{3\pi}(s)
\label{dom}
\end{eqnarray}
are respectively the inverse propagator  of the $\omega$ meson and its width
determined mainly by the  $\pi^+\pi^-\pi^0$ decay mode while
$W_{3\pi}(s)$ stands for the phase space volume of the final $3\pi$ state, see its expression Eq. (\ref{wdm2_3pi}) below.

The real part of the polarization operator of the  $\rho\omega$
transition is taken in the form
\begin{equation}
\mbox{Re}\Pi_{\rho\omega}=2m_\omega\delta_{\rho\omega}+
\frac{4\pi\alpha m^2_\rho m^2_\omega}{f_\rho f_\omega}(1/m^2_\omega-1/s);
\label{eq34}
\end{equation}
$\delta_{\rho\omega}$ is the amplitude of the  $\rho-\omega$ transition as
measured at the $\omega$ mass while the last term is aimed to take into
account the fast varying one photon contribution.

The expression for the imaginary part, $\mbox{Im}\Pi_{\rho\omega}$, is taken in the following way, different from Refs. \cite{argus,isoscalars}, see also \cite{ach92a}:
$$\mbox{Im}\Pi_{\rho\omega}=\sqrt{s}\bigg(\frac{g_{\phi K^+K^-}^2}{12\pi s}(q^3_{K^+K^-}-q^3_{K^0\bar K^0})\theta(s-4m^2_{K^0})+ $$
\begin{equation}
\frac{g^2_{\rho \omega\pi}}{12\pi s}(q^3_{K^{*+}K^-}-q^3_{K^{*0}\bar K^{*0}})\theta(s-4m^2_{K^{*0}})+\frac{g_{\rho\pi\gamma}g_{\omega\pi\gamma}}{3}q^3_{\pi\gamma}\bigg),\label{ImPi}
\end{equation}
\noindent representing $K\bar K,\ K^*K$ and $\pi\gamma$ intermediate states contributions. $\theta(s)$ is theta function, constants are obtained from the initial ones ($g_{\omega K^0\bar K^0}$, etc.) using SU(3) relations, see below. The constants $g_{V\pi\gamma}, V=\rho,\omega$, could be obtained via corresponding widths
\begin{equation}
\Gamma (V\to\pi\gamma)=\frac{g^2_{V\pi\gamma}}{96\pi m_V^3}(m_V^2-m_\pi^2).
\end{equation}

Note the effect of $\mbox{Im}\Pi_{\rho\omega}$ is small due to strong cancelation in the $K\bar K$ and $K^*K$ contributions.

\subsection{Isoscalar channels}
\label{isosc}

Processes $e^+e^-\to \rho + \rho'_1+\rho'_2+\omega + \omega'_1+\omega'_2+\phi+\phi_1'+\phi_2'\to K^+K^-, \pi^+\pi^-\pi^0$ are described basically under Ref. \cite{isoscalars}. In what follows we will often use also the notation $V_i$ ($i=1,2,3$) such that $V_1,V_2,V_3$ corresponds to $V,V^\prime_1,V^\prime_2$, and $V=\rho, \omega, \varphi$.

The cross section of the final state $f=\pi^+\pi^-\pi^0, K^+K^-$ production can be represented as
\begin{eqnarray}
\sigma_f&=&
{4\pi\alpha^2\over s^{3/2}}\bigg|\left(\frac{m^2_\rho}{f_\rho},
\frac{m^2_{\rho^{\prime}_1}}{f_{\rho^{\prime}_1}},
\frac{m^2_{\rho^{\prime}_2}}{f_{\rho^{\prime}_2}}\right)G^{-1}_\rho(s)
\left(
\begin{array}{c}
g_{\rho f}\\ g_{\rho^{\prime}_1 f}\\ g_{\rho^{\prime}_2 f}
\end{array}
\right)\;
                    \nonumber\\
& &+\left(\frac{m^2_\omega}{f_\omega},
\frac{m^2_{\omega^{\prime}_1}}{f_{\omega^{\prime}_1}},
\frac{m^2_{\omega^{\prime}_2}}{f_{\omega^{\prime}_2}},\frac{m^2_\varphi}{f_\varphi},\frac{m^2_{\varphi^{\prime}_1}}{f_{\varphi^{\prime}_1}},
\frac{m^2_{\varphi^{\prime}_2}}{f_{\varphi^{\prime}_2}}
\right)
G^{-1}_{\omega\varphi}(s)
\left(
\begin{array}{c}
g_{\omega f}\\ g_{\omega^{\prime}_1 f}\\ g_{\omega^{\prime}_2 f}\\g_{\varphi f}\\ g_{\varphi^{\prime}_1 f}\\ g_{\varphi^{\prime}_2 f}
\end{array}
\right)\;\bigg|^2P_f.
                    \nonumber\\
\label{eq2.1}
\end{eqnarray}

For the $\pi^+\pi^-\pi^0$ channel the first term describing $\rho_i$ contribution is zero. Remind the matrix $G_\rho(s)$ is shown in Eq. (\ref{eq7}), and the matrix $G_{\omega\varphi}(s)$ is
\begin{equation}
G_{\omega\varphi}(s)=\left(
\begin{array}{cccccc} D_\omega &-\Pi_{\omega\omega^{\prime}_1}&-\Pi_{\omega\omega^{\prime}_2}
& -\Pi_{\omega\varphi}&-\Pi_{\omega\varphi^{\prime}_1}&-\Pi_{\omega\varphi^{\prime}_2} \\
-\Pi_{\omega\omega^{\prime}_1} & D_{\omega^{\prime}_1} & -\Pi_{\omega^{\prime}_1\omega^{\prime}_2} & -\Pi_{\omega^{\prime}_1\varphi}&-\Pi_{\omega^{\prime}_1\varphi^{\prime}_1}&-\Pi_{\omega^{\prime}_1\varphi^{\prime}_2} \\
-\Pi_{\omega\omega^{\prime}_2} & -\Pi_{\omega^{\prime}_1\omega^{\prime}_2} & D_{\omega^{\prime}_2} & -\Pi_{\omega^{\prime}_2\varphi}&-\Pi_{\omega^{\prime}_2\varphi^{\prime}_1}&-\Pi_{\omega^{\prime}_2\varphi^{\prime}_2} \\
-\Pi_{\omega\varphi}&-\Pi_{\omega^{\prime}_1\varphi} & -\Pi_{\omega^{\prime}_2\varphi} & D_\varphi &-\Pi_{\varphi\varphi^{\prime}_1}&-\Pi_{\varphi\varphi^{\prime}_2}\\
-\Pi_{\omega\varphi^{\prime}_1}&-\Pi_{\omega^{\prime}_1\varphi^{\prime}_1} & -\Pi_{\omega^{\prime}_2\varphi^{\prime}_1} & -\Pi_{\varphi\varphi^{\prime}_1} &D_{\varphi^{\prime}_1}&-\Pi_{\varphi^{\prime}_1\varphi^{\prime}_2}\\
-\Pi_{\omega\varphi^{\prime}_2}&-\Pi_{\omega^{\prime}_1\varphi^{\prime}_2} & -\Pi_{\omega^{\prime}_2\varphi^{\prime}_2} & -\Pi_{\varphi\varphi^{\prime}_2} &-\Pi_{\varphi^{\prime}_1\varphi^{\prime}_2} & D_{\varphi^{\prime}_2}
\end{array}
\right)\;
\label{eq2.2}
\end{equation}
Note that in Ref. \cite{isoscalars} Okubo-Zweig-Iizuka (OZI) rule breaking transitions $\omega_i\leftrightarrow\varphi_j$, described by the terms $\Pi_{\omega_i\varphi_j}$, were taken into account under a special approximation to save machine time.

The factor $P_f$ for the  final states $f=\pi^+\pi^-\pi^0$, $K^+K^-$ reads, respectively,
\begin{equation}
P_f\equiv P_f(s)=
W_{3\pi}(\sqrt{s})\mbox{, } \frac{2q^3_{KK}}{3 s}\mbox{. }
\label{eq2.5}
\end{equation}

The factor
\begin{eqnarray}
W_{3\pi}(\sqrt{s})&=&{g^2_{\rho\pi\pi}\over12\pi^2}
\int_{2m_\pi}^{\sqrt{s}-m_\pi}dm q^3_\rho(m) q^3_\pi(m)
\nonumber\\
& &\times\int_{-1}^1dx(1-x^2)\left|{1\over D_\rho(m^2)}+
{1\over D_\rho(m^2_-)}+{1\over D_\rho(m^2_+)}\right|^2,
\label{wdm2_3pi}
\end{eqnarray}
where
\begin{eqnarray*}
m^2_\pm&=&{1\over2}(s+3m_\pi^2-m^2)      \nonumber\\
& &\pm{2\sqrt{s}\over m}q_\pi(m)q_\rho(m)x,   \nonumber\\
q_\rho(m)&=&q(\sqrt{s},m,m_\pi),           \nonumber\\
q_\pi(m)&=&q(m,m_\pi,m_\pi),
\end{eqnarray*}
is responsible for the phase space volume of the $\pi^+\pi^-\pi^0$ final state.

The inverse propagator $D_{V_i}$ of the vector meson $V_i=\omega_i, \varphi_i$ and the imaginary part of the
polarization operator $\Pi_{V_iV_j}$ are, respectively,
\begin{equation}
D_{V_i}(s)=m^2_{V_i}-s-i\sqrt{s}\Gamma_{V_i}(s)
\label{prop}
\end{equation}
and
\begin{eqnarray}
\mbox{Im}\Pi_{V_iV_j}(s)&=&\sqrt{s}\left(g_{V_i\rho\pi}g_{V_j\rho\pi}
P_{\pi^+\pi^-\pi^0}\right.        \nonumber\\
& &+2g_{V_iK^+K^-}g_{V_jK^+K^-}P_{K^+K^-}   \nonumber\\
& &\left.+4g_{V_iK^{\ast+} K}g_{V_jK^{\ast+}K^-}P_{K^0_SK^+\pi^-}
\right.        \nonumber\\
& &\left.+g_{V_iV_1\pi^+\pi^-}g_{V_jV\pi^+\pi^-}W_{VP\pi}\right),
\label{imag}
\end{eqnarray}

\noindent where $W_{VP\pi}\equiv W_{VP\pi}(\sqrt{s},m_{V_1},m_\pi)$. The factors $2$ and $4$ take into account all possible charge combinations. The width of the $V_i$ meson can be represented as usual:
\begin{equation}
\Gamma_{V_i}(s)=\mbox{Im}\Pi_{V_iV_i}(s)/\sqrt{s}.
\label{width}
\end{equation}

The $K^{\ast}K$ channel is taken into account with the help of the $q\bar q$ model relations:
\begin{eqnarray}
g_{\rho_{1,2,3}K^{\ast+}\bar K^-}&=&{1\over2}g_{\omega_{1,2,3}\rho\pi},
\nonumber\\
g_{\omega_{1,2,3}K^{\ast+}\bar K^-}&=&{1\over2}g_{\omega_{1,2,3}\rho\pi},
\nonumber\\
g_{\varphi_{1,2,3}K^{\ast+}\bar K^-}&=&{1\over\sqrt{2}}
g_{\omega_{1,2,3}\rho\pi},
\label{relations}
\end{eqnarray}
and the SU(2) related to them.

We also use the relations
\begin{eqnarray}
g_{\rho_{1,2,3}K^+K^-}&=&-{1\over\sqrt{2}}g_{\varphi_{1,2,3}K^+K^-}=-g_{\rho_{1,2,3}K^0\bar K^0},
\nonumber\\
g_{\omega_{1,2,3}K^+K^-}&=&-{1\over\sqrt{2}}g_{\varphi_{1,2,3}K^+K^-}=g_{\omega_{1,2,3}K^0\bar K^0},
\nonumber\\
f_{\omega_{1,2,3}}&=&3f_{\rho_{1,2,3}},
\nonumber\\
f_{\phi_{1,2,3}}&=&\frac{-3}{\sqrt{2}}f_{\rho_{1,2,3}},
\label{su3pred}
\end{eqnarray}
but don't demand their strict realization, see Sec. \ref{dataDescr}.

The expressions for the partial widths could include the energy-dependent
factors $C_f(s)$ which,
analogously to the well known Blatt-Weiskopf centrifugal factors, are
aimed at  restricting a too fast growth of the partial widths with the energy
rise. They are somewhat arbitrary under the demand of $\sqrt{s}\Gamma(s)\to$
const at $\sqrt{s}\to\infty$. In practice, the only mode with a strong
dependence is
the vector (V)+pseudoscalar (P) one,  and our choice for the factor
multiplying corresponding coupling constant is
\begin{equation}
C_{VP}(s)={1+(R_{VP}m_0)^2\over1+(R_{VP}\sqrt{s})^2},
\label{blatt}
\end{equation}
where $m_0$ is the mass of the resonance and $R_{VP}$ is the so-called range
parameter.

\begin{figure}[h]
\begin{center}
\begin{tabular}{ccc}
\includegraphics[width=8cm]{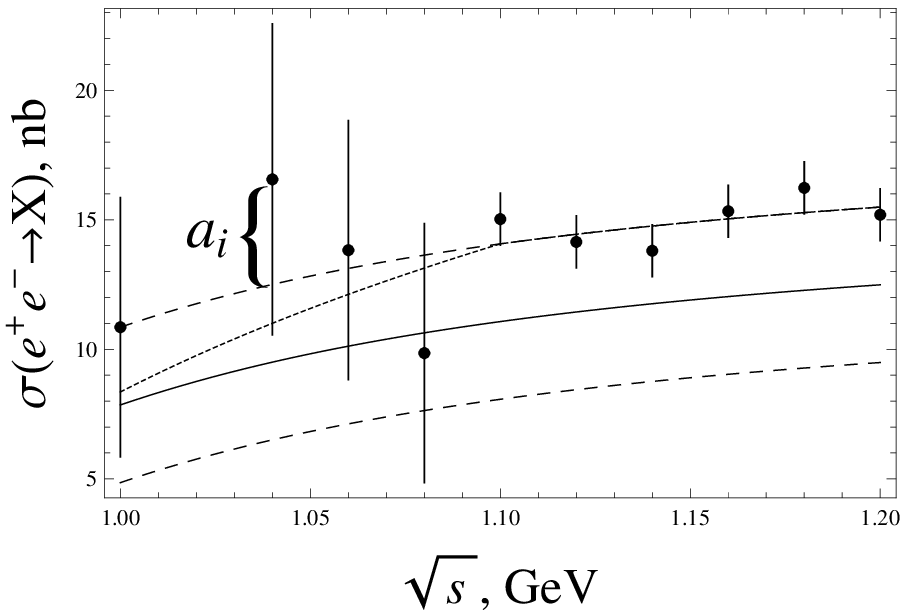}& \includegraphics[width=8cm]{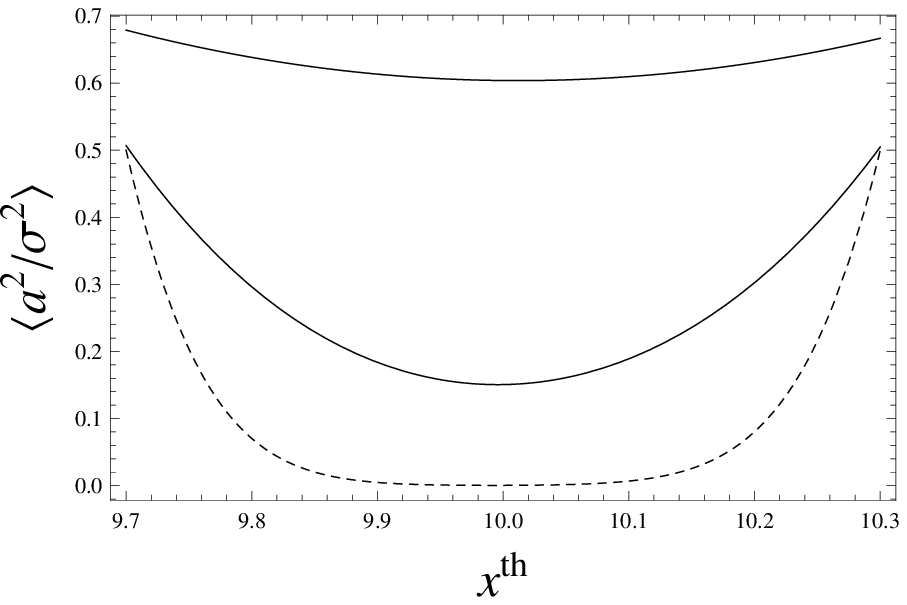}\\
a) & b)
\end{tabular}
\end{center}
\caption{a) A fictional situation for an illustration: cross section of some process $e^+e^-\to X$ in the energy region (1 GeV, 1.2 GeV). Solid line is a theoretical curve, dashed lines are borders, dotted line is a true cross section, $a_i$ is the distance to the corridor, see Eq. (\ref{eqa}), points are experimental data.\\
b) Average value $\langle a^2/\sigma^2\rangle$ as a function of $x^{th}$ with $\beta=0.03$, $x^0=10$. The upper solid line is for $\sigma=0.1$, the bottom solid line is for $\sigma=0.3$, and the dashed line is for $\sigma=1$. The shown $x^{th}$ region is from $(1-\beta) x^0=9.7$ up to $(1+\beta)x^0=10.3$. }
\label{GFunc}
\end{figure}

\section{Modified $\chi^2$ function}
\label{SecChi2}

For each i-th experimental point there is a true value of a cross section $x^0_i$, model prediction $x^{th}_i$, experimental result $x^{exp}_i$, and experimental error $\sigma_i$. Since our model is not completely precise, we introduce a maximally allowed relative model error $\beta$, and our hypothesis to check is that $|x^{th}_i-x^0_i| \leq \beta x^0_i$ in every experimental point. For example, $\beta=0.1$ means that accuracy of the model is 10\%.

To illustrate these definitions lets imagine the fictional situation shown in Fig. \ref{GFunc}a. There is a cross section of some process, theoretical curve - its values in experimental energies are our $x^{th}_i$, the true curve - its values are $x^0_i$, the borders limiting the true curve are the theoretical curve divided by $(1+\beta)$ and $(1-\beta)$ (the condition $|x^{th}_i-x^0_i| \leq \beta x^0_i$ means that $\frac{x^{th}_i}{1+\beta}\leq x^0_i \leq \frac{x^{th}_i}{1-\beta}$). Experimental points in Fig. \ref{GFunc}a are our $x^{exp}_i$, experimental errors are our $\sigma_i$.

In the familiar case with $\beta=0$ usual $\chi^2$ criterion is commonly used:
\begin{equation}\chi^2=\sum_i (x^{exp}_i-x^{th}_i)^2/\sigma_i^2\,,\end{equation}
and by the obtained $\chi^2$ value one finds out the level of hypothesis validity.

When $\beta > 0$ it can seem adequate at first glance to form an analog of usual $\chi^2$ function, $\bar \chi^2$, in such a way: if for i-th point $x^{exp}_i$ is inside the corridor $\Big(\frac{x^{th}_i}{1+\beta}, \frac{x^{th}_i}{1-\beta}\Big)$, the difference $|x^{exp}_i-x^{th}_i|$ is fully blamed to model inaccuracy, so the contribution from this point to $\bar \chi^2$ is zero. If $x^{exp}_i$ is outside the corridor, the distance to the corridor, $a_i$, is taken into consideration, and the contribution to $\bar \chi^2$ is $\frac{a_i^2}{\sigma_i^2}$. Visually $a_i$ is shown in Fig. \ref{GFunc}a for the point at $\sqrt{s}=1.04$ GeV. Formally,
\begin{equation}a_i=\theta \Big(x^{exp}_i-\frac{x^{th}_i}{1-\beta}\Big)\,\Big(x^{exp}_i-\frac{x^{th}_i}{1-\beta}\Big)+\theta \Big(\frac{x^{th}_i}{1+\beta}-x^{exp}_i\Big)\,\Big(x^{exp}_i-\frac{x^{th}_i}{1+\beta}\Big)\,,\label{eqa}\end{equation}
where $\theta(x)$ is the $\theta$-function. With this definition $a_i$ is negative when $x^{exp}_i$ is below than bottom limit of the corridor, it doesn't matter for our purposes.

But such function $\bar \chi^2$ does not have the same features as usual $\chi^2$ function. It would be incorrect to use the obtained $\bar \chi^2$ value as usual $\chi^2$ value to determine the level of hypothesis validity. Even average value of $\bar \chi^2$, $\langle \bar \chi^2\rangle$, differs a lot from $\langle \chi^2\rangle$ in general case because $\langle a_i^2 \rangle$ differs from $\sigma_i^2$.

What can be done in this situation? Since our experimental data consist of very many experimental points $N_p$ and degrees of freedom $N_{df}$, while number of free parameters is much less than $N_{df}$, we modify $\bar \chi^2$ and estimate
\begin{equation}\bar \chi^2=\sum_{i} \frac{1}{\langle a_i^2/\sigma_i^2\rangle}\,\frac{a^2_i}{\sigma_i^2}=\sum_{i} \frac{a^2_i}{\langle a_i^2\rangle}\approx N_{df}\,. \label{realSum}\end{equation}
So in general we should find model accuracy that fulfills Eq. (\ref{realSum}), this should provide a good estimation. To do this we need to estimate $\langle a_i^2\rangle$.

Let us consider some experimental point, where there is a true result $x^0$ (index "i" is omitted for convenience) and theoretical prediction $x^{th}$. Experimental result $x^{exp}$ is a random value with probability density $f_{exp}(x^{exp})$, usually normal distribution is assumed:
\begin{equation} f_{exp}(x)=\frac{1}{\sqrt{2\pi}\sigma}e^{-\frac{(x-x^0)^2}{2\sigma^2}}\,. \label{fGauss}\end{equation}

For $a\equiv a_i$ we have probability density
\begin{equation} f_a(a)=K\delta(a)+\theta(a)f_{exp}\Big(a+\frac{x^{th}}{1-\beta}\Big) + \theta(-a)f_{exp}\Big(a+\frac{x^{th}}{1+\beta})\,,\label{fa} \end{equation}
where
\begin{equation} K=\int^{x^{th}/(1-\beta)}_{x^{th}/(1+\beta)}f_{exp}(x^{exp})\,dx^{exp}\,.\label{K}\end{equation}

As usual,
\begin{equation}\int^\infty_{-\infty} f_a(a)\,da=1\,.\end{equation}

To calculate the average value of $a^2/\sigma^2$ we use Eq. (\ref{fGauss}): $$\Big\langle\frac{a^2}{\sigma^2}\Big\rangle=\int^{\infty}_{-\infty}\,da\,\frac{a^2}{\sigma^2}f_a(a)=$$
$$(d_{-}^2+\frac{1}{2})\,(1-Erf(d_{-}))-\frac{d_{-}}{\sqrt{\pi}}e^{-d_{-}^2}+(d_{+}^2+
\frac{1}{2})\,(1+Erf(d_{+}))+\frac{d_{+}}{\sqrt{\pi}}e^{-d_{+}^2}\,,$$
\begin{equation} Erf(x)=\frac{2}{\sqrt{\pi}}\int^x_0 e^{-t^2}\,dt\,,\ d_{+}=\frac{\frac{x^{th}}{1+\beta}-x^0}{\sqrt{2}\sigma}\,,\  d_{-}=\frac{\frac{x^{th}}{1-\beta}-x^0}{\sqrt{2}\sigma}\,.\label{midValFull} \end{equation}

If $x^{th}=x^0$ and $\beta=0$ we get $\langle a^2/\sigma^2\rangle=1$. In case close to this situation, when $|x^{th}-x^0|<<\sigma$ and $\beta x^0 <<\sigma$, $\langle a^2/\sigma^2\rangle$ is close to 1. But in general case $\langle a^2/\sigma^2\rangle$ is significantly less than $1$ and can be even close to $0$, see Fig. \ref{GFunc}b. That is why treating $\bar \chi^2$ value as the $\chi^2$ value in $\chi^2$ criterion would lead to overestimation of the confidence level of the hypothesis.

In the borderline case $x^{th}=(1\pm\beta)x^0$ and small $\sigma$ we have $\langle a^2/\sigma^2\rangle\approx 0.5$. This result can be explained with the help of Fig. \ref{GFunc}a, see the situation in the energy region from $1.1$ GeV up to $1.2$ GeV. Here the true cross section coincides with the upper border line of the corridor, and experimental points randomly fluctuate above and below the border line, so half of them gives zero contribution to $\bar \chi^2$, and the second half gives $1$ in average - the contribution to usual $\chi^2$ in the usual situation, since the model error is precisely subtracted in this case. So we have twice as small contribution in this situation in comparison with the naive $\bar \chi^2$ definition discussed in the beginning of this section.

This visual effect does motivate such an excursion into statistics. Note that the effect of halving takes place not only for the distribution Eq. (\ref{fGauss}), but for any symmetrical distribution.

To calculate $\langle a_i^2/\sigma_i^2\rangle$ and $\bar \chi^2$ via Eq. (\ref{midValFull}) we estimate true value $x^0_i=x^{th}_i/(1-\beta)$. On the one hand, it leads to "masking effect"\ -- underestimation of model error because of the fact that in some points $x^0_i$ can be closer to $x^{th}_i$ than $x^{th}_i/(1-\beta)$, in such cases $\langle a_i^2/\sigma_i^2\rangle$ are less than our estimation. One can see this situation in Fig. \ref{GFunc}a in the energy region from $1$ GeV up to $1.08$ GeV.

On the other hand, in reality we don't know true values of cross sections. Our hypothesis gives us freedom in choice of model error inside the corridor, and we do not know a particular situation making suspect "masking effect" stronger than in another particular situation. This suspicion is general: it is hard to believe that model error always ideally follows experimental data.

So we take into account the effect of halving, but ignore "masking effect". By the way, estimation Eq. (\ref{realSum}) can be treated as a kind of compensation because it corresponds to confidence level close to $50\%$, which is more strict than $10\%$ that is widely accepted. It is an essential factor. In case of a precise model without free parameters checked in $1000$ experimental points usual $\chi^2=\sum_{i} (x^{exp}_i-x^0_i)^2/\sigma_i^2$ has the average value $1000$ and, in Gaussian case, the standard deviation $\sqrt{2\cdot 1000}\approx 45$, or $4.5\%$ of the average value. In the given paper we have less than $1000$ experimental points (and even fewer degrees of freedom), so this factor is even more valuable.

Note that usual consideration (with a precise model and usual $\chi^2$) is also approximate from the very beginning starting from the facts that in reality distributions are not Gaussian, errors $\sigma_i$ provided by experimentalists are only estimations of real errors, and systematic errors are correlated. We think that additional assumptions made above do not change the level of assumptions. In general, our results below could be formulated like this: statistically the data is in agreement with the hypothesis that in each experimental point with $|a_i|>0$ the $x^{th}_i$ value deviates from the true value $x^0_i$ by $\beta\cdot 100$ percent. In the points where $a_i=0$ the $x^{th}_i$ values are closer to $x^0_i$.

\section{The data description}
\label{dataDescr}
We follow the scenario based on the quark model with coupling constants approximately obeying SU(3) relations, mass hierarchy, and masses of all resonances not far from the Ref. \cite{pdg-2020} values. To provide these requirements we set some limits on parameters and their combinations. For example, we demand $m_{\varphi^{\prime}_2}\geq$2 GeV and $m_{\varphi^{\prime}_1}- m_{\omega^{\prime}_1}\geq$150 MeV, and fitting procedure chooses to satisfy these limits on the border, $m_{\varphi^{\prime}_2}=$2 GeV and $m_{\varphi^{\prime}_1}- m_{\omega^{\prime}_1}=$150 MeV, see Table I.


Technically the function to minimize is
\begin{eqnarray}
\bar \chi^2_{tot}=\bar \chi^2+\bar \chi^2_{add}\,, \nonumber\\
\bar \chi^2=\bar \chi^2_{\pi\pi}+\bar \chi^2_{\omega\pi}+\bar \chi^2_{\eta\pi\pi}+\bar \chi^2_{KK}+\bar \chi^2_{3\pi}\,,
\end{eqnarray}
where $\bar \chi^2_{\pi\pi}$ etc. are modified $\chi^2$ functions for all five processes, and $\bar \chi^2_{add}$ provides strict limits and being close to approximate predictions.

On the first step we add to $\bar \chi^2_{add}$ terms like \begin{equation}\frac{(2\mbox{ GeV}-m_{\varphi^{\prime}_2} + |m_{\varphi^{\prime}_2}-2\mbox{ GeV}|)^2}{\varepsilon^2},\end{equation}
where $\varepsilon$ is small enough. This term vanishes when the requirement $m_{\varphi^{\prime}_2}\geq$2 GeV is satisfied, so the parameters ($m_{\varphi^{\prime}_2}$ in this case) don't suffer additional pressure. But if the limit is violated and $m_{\varphi^{\prime}_2}$ goes below 2 GeV, this term becomes large, preventing essential violation of the limit.

On the second step we collect violated limits and set corresponding parameters to limit values directly ($m_{\varphi^{\prime}_2}=2\mbox{ GeV}$ or $m_{\varphi_1}=m_{\omega_1}+$150 MeV in cases above). On the third step, after obtaining $\bar \chi^2_{tot}$ minimum, we free again these parameters and check that they again try to violate limits.

On the first step we have the following violated parameters and combinations: $|m_{\rho_1}-m_{\omega_1}|\leq$30 MeV, $m_{\rho_2}\leq$1.75 GeV, $m_{\omega_2}\leq$1.75 GeV, $m_{\varphi_2}\geq$2 GeV, $m_{\varphi_1}- m_{\omega_1}\geq$150 MeV, $|g_{\phi_1\rho\pi}/g_{\omega_1\rho\pi}| \leq$0.1, $\mbox{Re}\Pi_{\phi_1\phi_2}\leq$0.9 GeV$^2$ (the last is for a kind of safety, too large $\mbox{Re}\Pi_{\phi_1\phi_2}$ raises the question about its energy dependence, we plan to consider it later with the help of Ref. \cite{AchKozh-2013}). All of them are set to border values or close to them, see Table I, the number of free parameters of the fit is correspondingly reduced.

Besides, some parameters and combinations are not strictly limited, but pushed to be "as close to some value, as possible, but not by a dear price". For example, we require $g_{\varphi^{\prime}_1 K^+K^-}/g_{\rho^{\prime}_1 K^+K^-}$ to be not far from $-\sqrt{2}$ (SU(3) prediction) by adding to $\bar \chi^2_{add}$ the term \begin{equation}\frac{\bigg(\frac{g_{\varphi^{\prime}_1 K^+K^-}}{g_{\rho^{\prime}_1 K^+K^-}}+\sqrt{2}\bigg)^2}{0.1^2},\end{equation}

\noindent which also reduces the number of free parameters. We add such terms for SU(3) predictions Eqs. (\ref{su3pred}) and $y_{\eta i}$ (remind Eq. (\ref{yDef})). Besides, we set $m_\phi=1019.461$ according to Ref. \cite{pdg-2020}.

Speaking about number of degrees of freedom (n.d.f.), we have 322 points on the $e^+e^-\to\pi^+\pi^-$ reaction below 2 GeV \cite{babarPP}, 56 points on $e^+e^-\to\omega\pi^0$ \cite{sndOmPi}, 39 points on $e^+e^-\to\eta\pi^+\pi^-$ \cite{sndEtaPP}, 80 points on $e^+e^-\to K^+K^-$ \cite{KK} and 156 points on $e^+e^-\to\pi^+\pi^-\pi^0$ \cite{3Pi}. Since we need to describe well every reaction, it is not enough to require Eq. (\ref{realSum}) for all 653 points. We demand $\bar \chi^2$ for each decay ($\bar \chi^2_{\pi\pi}$ etc.) to be less or insignificantly higher than n.d.f. corresponding to this decay.

It has turned out that the easiest reactions to describe with our model are $e^+e^-\to\pi^+\pi^-$ and $e^+e^-\to\omega\pi^0$, then $e^+e^-\to\pi^+\pi^-\pi^0$, then $e^+e^-\to K^+K^-$, and the $e^+e^-\to\eta\pi^+\pi^-$ reaction is most problematic. So we refer maximal number of residual free parameters dealing with isovector mesons to the $e^+e^-\to\pi^+\pi^-$ reaction, and maximal number of free parameters dealing with isoscalar mesons to the $e^+e^-\to\pi^+\pi^-\pi^0$ reaction. As a result we refer 15 residual free parameters to the $e^+e^-\to\pi^+\pi^-$ process, 3 to $e^+e^-\to\omega\pi^0$, 0 to $e^+e^-\to\eta\pi^+\pi^-$, 2 to $e^+e^-\to K^+K^-$, and 14 to $e^+e^-\to\pi^+\pi^-\pi^0$.

So we demand $\bar \chi^2_{\pi\pi}$ to be less or insignificantly higher than 307, $\bar \chi^2_{\omega\pi}$ -- 53, $\bar \chi^2_{\eta\pi\pi}$ -- 39, $\bar \chi^2_{KK}$ -- 78, and $\bar \chi^2_{3\pi}$ -- 142.

\begin{figure}[h]
\begin{center}
\includegraphics[width=18cm]{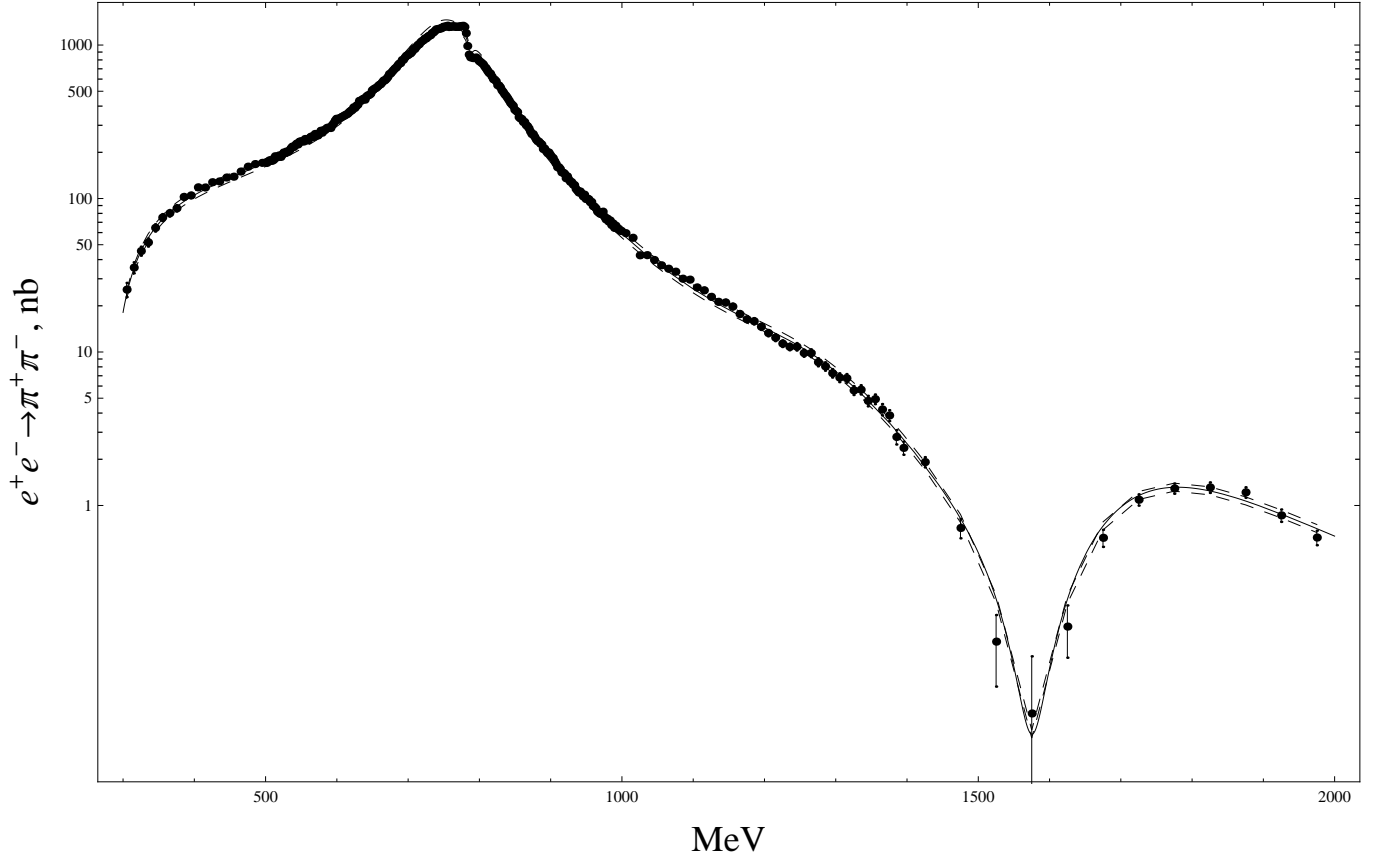}
\end{center}
\caption{$\sigma(e^+e^-\to\pi^+\pi^-)$, nb. Solid line is for Fit 1, dashed lines are borders, the data are from Ref. \cite{babarPP}.}
\label{PPfig}
\end{figure}

\begin{figure}[h]
\begin{center}
\begin{tabular}{ccc}
\includegraphics[width=8cm]{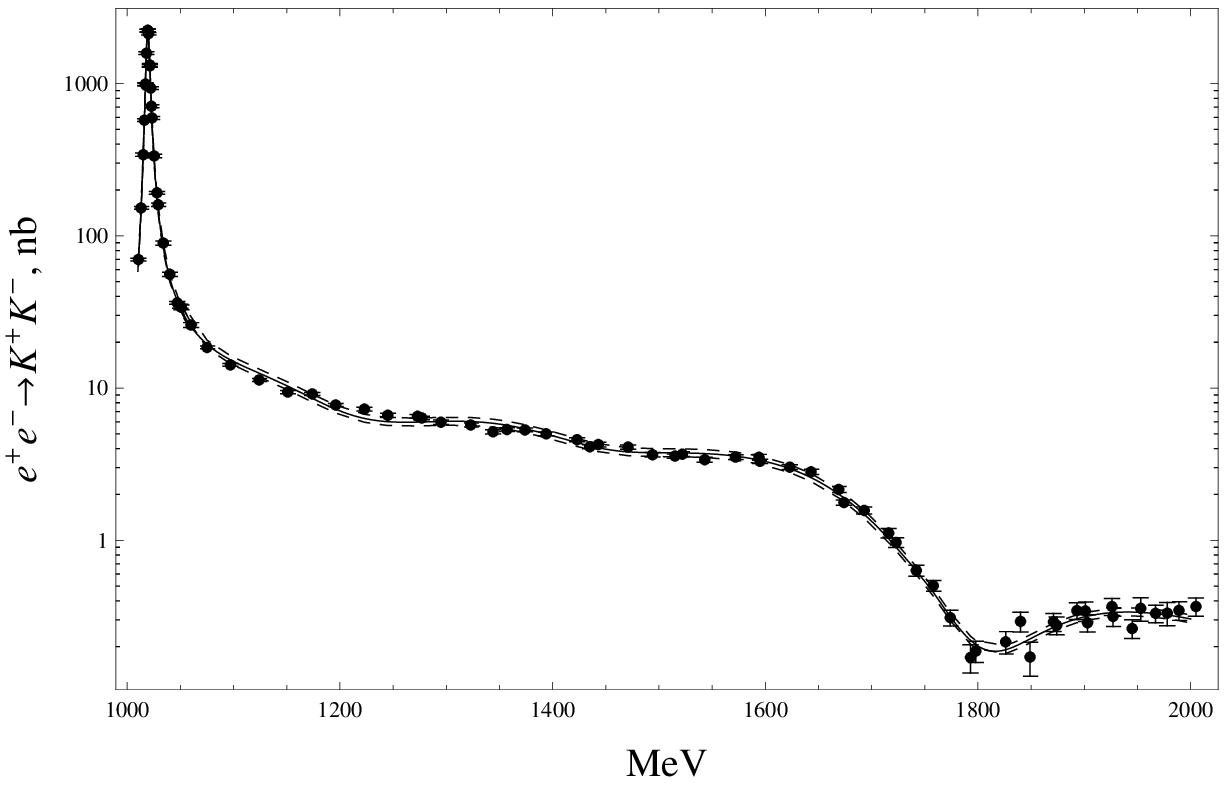}& \includegraphics[width=8cm]{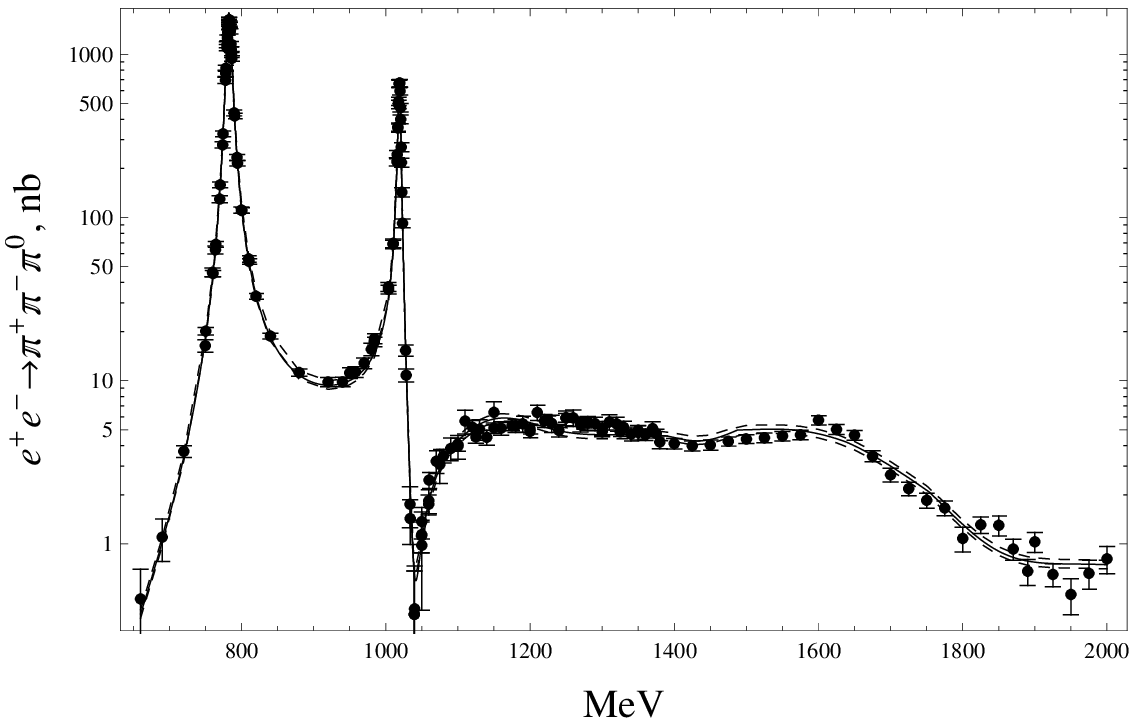}
\end{tabular}
\end{center}
\caption{$\sigma(e^+e^-\to K^+K^-)$ and $\sigma(e^+e^-\to\pi^+\pi^-\pi^0)$, nb. Solid lines are for Fit 1, dashed lines are borders, the data are from Refs. \cite{KK} and \cite{3Pi} correspondingly.}
\label{KK3Pfig}
\end{figure}

With all this we obtain Fit 1 in Table I and overall model accuracy $\beta=0.06$. Figs. \ref{PPfig}-\ref{KK3Pfig} show that the data is described well in all energy regions with given accuracy.

\begin{figure}[h]
\begin{center}
\begin{tabular}{ccc}
\includegraphics[width=8cm]{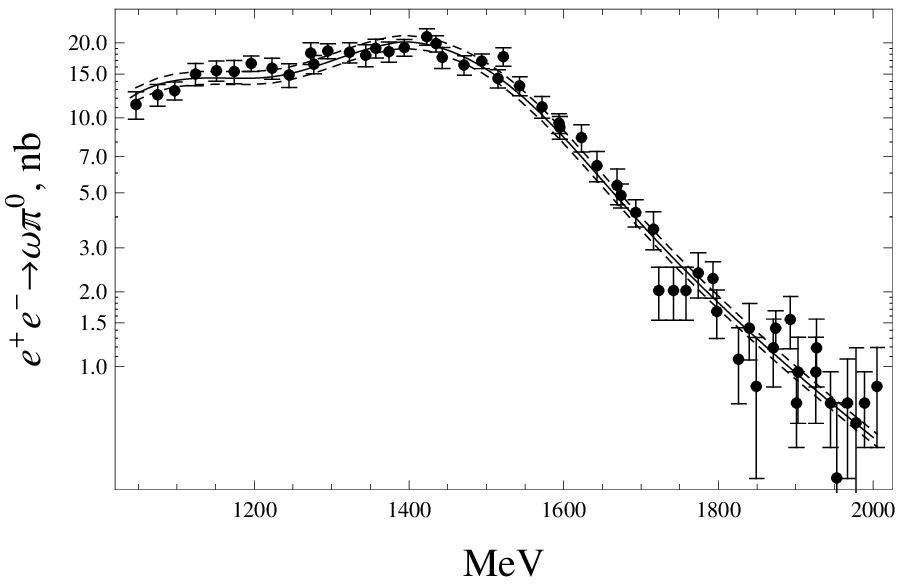}& \includegraphics[width=8cm]{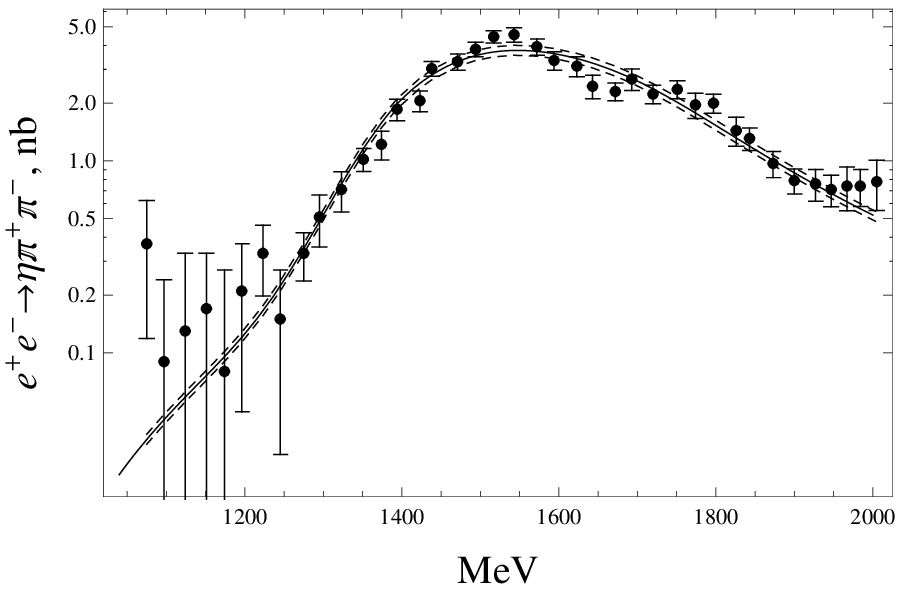}
\end{tabular}
\end{center}
\caption{$\sigma(e^+e^-\to\omega\pi^0)$ and $\sigma(e^+e^-\to\eta\pi^+\pi^-)$, nb. Solid lines are for Fit 1, dashed lines are borders, the data are from Refs. \cite{sndOmPi} and \cite{sndEtaPP} correspondingly.}
\end{figure}

An essential question deals with $m_\phi$. The mass provided by Ref. \cite{pdg-2020} is some average of peak positions in some reactions, these positions are determined by different contributions, including those unrelated to bare or physical $\phi$ at all. But approximately we treat the mass in Ref. \cite{pdg-2020} as a physical mass, formed by mixing of bare $\phi$ meson with primarily $\varphi^{\prime}_1$ and $\varphi^{\prime}_2$ and secondarily with other states due to common decay channels. This picture does not contradict the small width of the $\phi$ meson and $s\bar s$ dominance in $\phi$, see Sec. \ref{notePhi} for details.

Making $m_\phi$ a free parameter gives $m_\phi\approx 1028.3$ MeV, see Fit 2. $\bar \chi^2$ is significantly improved in this case.

It turns out that isovector channels $e^+e^-\to\pi^+\pi^-$ and $e^+e^-\to\omega\pi^0$ can be described much better than with $\beta=0.06$ practically without prejudice to other channels description. Simultaneous data description with $\beta_{\pi\pi}=\beta_{\omega\pi}=0$, $\beta_{KK}=\beta_{3\pi}=\beta_{\eta\pi\pi}=0.06$ gives $\bar \chi^2_{\pi\pi}=222$, $\bar \chi^2_{\omega\pi}=44$, $\bar \chi^2_{\eta\pi\pi}=40$, $\bar \chi^2_{KK}=76$, and $\bar \chi^2_{3\pi}=135$ (in this fitting the $\phi$ mass was kept on Ref. \cite{pdg-2020} value).

Note that the data is sensitive enough to change of overall accuracy. For example, $\beta=0.05$ gives $\bar \chi^2_{3\pi}=151$, total $\bar \chi^2=368$ (the $\phi$ mass is also on Ref. \cite{pdg-2020} value). We can conclude that on the current stage isoscalar channels determine the overall accuracy.

\section{A note on $m_\phi$ and radiative $\phi$ decays}
\label{notePhi}
The society get used to the fact that $\phi(1020)$ is a narrow resonance with mass close to $1020$ MeV. How does it agree with our Fit 2 with $m_\phi=1028.3$ MeV?

The point is that in Table I the mass of bare $\phi$ meson is shown. In GVDM, after mixing with other vector states due to common decay channels, we have the physical state with mass close to $1020$ MeV, i.e. the familiar situation. The main effect comes from $\phi$ mixing with $\phi^{\prime}_{1,2}$. As far as we know, this picture does not contradict anything.

\begin{center}
Table I. Properties of the resonances and main characteristics\\
\begin{tabular}{|c|c|c|c|c|c|}\hline

Fit & 1 & 2 \\ \hline

$m_\rho$, MeV & $776.43$ & $769.52$ \\ \hline

$g_{\rho\pi\pi}$ & $6.145$ & $5.948$ \\ \hline

$g_{\rho K^+K^-}$ & $-2.356$ & $-2.059$ \\ \hline

$g_{\rho \omega \pi}\mbox{, GeV}^{-1}$ & $16.000$ & $16.015$ \\ \hline

$g_{\rho\rho^0\pi^+\pi^-}$ & $58.405$ & $64.180$ \\ \hline

$g_{\rho\rho^+\rho^-}$ & $55.924$ & $55.994$ \\ \hline

$f_\rho$ & $5.170$ & $5.567$ \\ \hline

$m_{\rho^{\prime}_1}\mbox{, MeV}$ & $1454.65$ & $1450.59$ \\ \hline

$g_{\rho^{\prime}_1\pi^+\pi^-}$ & $-3.568$ & $-4.057$ \\ \hline

$g_{\rho^{\prime}_1 K^+K^-}$ & $5.529$ & $5.332$ \\ \hline

$g_{\rho^{\prime}_1\omega\pi}\mbox{, GeV}^{-1}$ & $9.390$ & $8.415$ \\ \hline

$g_{\rho^{\prime}_1\rho^0\pi^+\pi^-}$ & $-44.035$ & $-43.370$ \\ \hline

$g_{\rho^\prime_1\rho^+\rho^-}$ & $6.429$ & $5.238$ \\ \hline

$f_{\rho^\prime_1}$ & $2.163$ & $2.363$ \\ \hline

$m_{\rho^{\prime}_2}\mbox{, MeV}$ & $1760.15$ & $1760.09$ \\ \hline

$g_{\rho^{\prime}_2\pi^+\pi^-}$ & $1.558$ & $1.805$ \\ \hline

$g_{\rho^{\prime}_2 K^+K^-}$ & $-3.259$ & $-3.156$ \\ \hline

$g_{\rho^{\prime}_2\omega\pi}\mbox{, GeV}^{-1}$& $-5.793$ & $-5.313$ \\ \hline

$g_{\rho^{\prime}_2\rho^0\pi^+\pi^-}$ & $-117.99$ & $-115.76$ \\ \hline

$g_{\rho^\prime_2\rho^+\rho^-}$ & $-3.237$ & $-2.651$ \\ \hline

$f_{\rho^\prime_2}$ & $1.464$ & $1.588$ \\ \hline

Re$\Pi_{\rho\rho^\prime_1}$, GeV$^2$ & $0.124244$ & $0.065537$ \\ \hline

Re$\Pi_{\rho\rho^\prime_2}$, GeV$^2$ & $-0.059230$ & $-0.015155$ \\ \hline

Re$\Pi_{\rho^\prime_1\rho^\prime_2}$, GeV$^2$ & $-0.501113$ & $-0.500960$ \\ \hline

\end{tabular}
\end{center}

\vspace{15mm}

\begin{center}
Table I. Properties of the resonances and main characteristics (continuation)\\
\begin{tabular}{|c|c|c|c|c|c|}\hline

Fit & 1 & 2 \\ \hline

$m_\omega$, MeV & $787.19$ & $785.93$ \\ \hline

$g_{\omega K^+K^-}$ & $-2.356$ & $-2.059$ \\ \hline

$f_\omega$ & $14.469$ & $15.713$ \\ \hline

$m_{\omega^\prime_1}\mbox{, MeV}$ & $1488.00$ & $1485.37$ \\ \hline

$g_{\omega^\prime_1\rho\pi}\mbox{, GeV}^{-1}$ & $-15.036$ & $-15.442$\\ \hline

$g_{\omega^\prime_1 K^+K^-}$ & $5.614$ & $5.427$ \\ \hline

$g_{\omega^\prime_1\omega\pi^+\pi^-}$ & $133.192$ & $137.541$ \\ \hline

$f_{\omega^\prime_1}$ & $7.176$ & $8.123$ \\ \hline

$m_{\omega^\prime_2}\mbox{, MeV}$ & $1753.43$ & $1751.96$ \\ \hline

$g_{\omega^\prime_2\rho\pi}\mbox{, GeV}^{-1}$ & $7.271$ & $7.444$\\ \hline

$g_{\omega^\prime_2 K^+K^-}$ & $-2.141$ & $-1.942$ \\ \hline

$g_{\omega^\prime_2\omega\pi^+\pi^-}$ & $126.376$ & $127.965$ \\ \hline

$f_{\omega^\prime_2}$ & $3.748$ & $4.075$ \\ \hline

Re$\Pi_{\omega\omega^\prime_1}$, GeV$^2$ & $0.022788$ & $0.018400$ \\ \hline

Re$\Pi_{\omega\omega^\prime_2}$, GeV$^2$ & $-0.039507$ & $-0.027154$ \\ \hline

Re$\Pi_{\omega^\prime_1\omega^\prime_2}$, GeV$^2$ & $-0.294809$ & $-0.260907$ \\ \hline

$m_\varphi$, MeV & $1019.461$ & $1028.283$ \\ \hline

$g_{\varphi K^+K^-}$ & $4.010$ & $3.090$ \\ \hline

$g_{\varphi\rho\pi}\mbox{, GeV}^{-1}$ & $0.746$ & $0.853$\\ \hline

$f_\varphi$ & $-12.336$ & $-13.229$ \\ \hline

$m_{\varphi^\prime_1}\mbox{, MeV}$ & $1638.00$ & $1635.37$ \\ \hline

$g_{\varphi^\prime_1 K^+K^-}$ & $-7.957$ & $-8.031$ \\ \hline

$g_{\varphi^\prime_1\rho\pi}\mbox{, GeV}^{-1}$ & $1.504$ & $1.544$\\ \hline

$f_{\varphi^\prime_1}$ & $-4.092$ & $-4.640$ \\ \hline

$m_{\varphi^\prime_2}\mbox{, MeV}$ & $2000$ & $2000$ \\ \hline

$g_{\varphi^\prime_2 K^+K^-}$ & $3.379$ & $3.219$ \\ \hline

$g_{\varphi^\prime_2\rho\pi}\mbox{, GeV}^{-1}$ & $0.767$ & $0.762$\\ \hline

$f_{\varphi^\prime_2}$ & $-3.554$ & $-3.929$ \\ \hline

\end{tabular}
\end{center}

\vspace{5mm}

\begin{center}
Table I. Properties of the resonances and main characteristics (continuation)\\
\begin{tabular}{|c|c|c|c|c|c|}\hline

Fit & 1 & 2 \\ \hline

Re$\Pi_{\varphi\varphi^\prime_1}$, GeV$^2$ & $-0.009294$ & $-0.112439$ \\ \hline

Re$\Pi_{\varphi\varphi^\prime_2}$, GeV$^2$ & $0.010112$ & $0.098679$ \\ \hline

Re$\Pi_{\varphi^\prime_1\varphi^\prime_2}$, GeV$^2$ & $-0.9$ & $-0.9$ \\ \hline

$y_{\eta 1}$ & $1.264$ & $1.320$\\ \hline

$y_{\eta 2}$ & $1.178$ & $1.212$\\ \hline

$R_{\omega\pi^0}\mbox{, GeV}^{-1}$ & $0.001$ & $0.000$
\\ \hline

$R_{\rho\pi^0}\mbox{, GeV}^{-1}$ & $0.502$ & $0.507$
\\ \hline

$3f_\rho/f_\omega$ & $1.07$ & $1.06$\\ \hline

$3f_{\rho^\prime_1}/f_{\omega^\prime_1}$ & $0.90$ & $0.87$\\ \hline

$3f_{\rho^\prime_2}/f_{\omega^\prime_2}$ & $1.17$ & $1.17$\\ \hline

$-3f_\rho/(\sqrt{2}f_\varphi$) & $0.89$ & $0.89$\\ \hline

$-3f_{\rho^\prime_1}/(\sqrt{2}f_{\varphi^\prime_1}$) & $1.12$ & $1.08$\\ \hline

$-3f_{\rho^\prime_2}/(\sqrt{2}f_{\varphi^\prime_2}$) & $0.87$ & $0.86$\\ \hline

$g_{\rho K^+K^-}/g_{\omega K^+K^-}$ & $1$ & $1$\\ \hline

$g_{\rho^\prime_1 K^+K^-}/g_{\omega^\prime_1 K^+K^-}$ & $0.98$ & $0.98$\\ \hline

$g_{\rho^\prime_2 K^+K^-}/g_{\omega^\prime_2 K^+K^-}$ & $1.52$ & $1.63$\\ \hline

$-\sqrt{2}g_{\rho K^+K^-}/g_{\varphi K^+K^-}$ & $0.83$ & $0.94$\\ \hline

$-\sqrt{2}g_{\rho^\prime_1 K^+K^-}/g_{\varphi^\prime_1 K^+K^-}$ & $0.98$ & $0.94$\\ \hline

$-\sqrt{2}g_{\rho^\prime_2 K^+K^-}/g_{\varphi^\prime_2 K^+K^-}$ & $1.36$ & $1.39$\\ \hline

$g_{\varphi\rho\pi}/g_{\omega\rho\pi}$ & $0.047$ & $0.053$\\ \hline

$g_{\varphi^\prime_1\rho\pi}/g_{\omega^\prime_1\rho\pi}$ & $-0.1$ & $-0.1$\\ \hline

$g_{\varphi^\prime_2\rho\pi}/g_{\omega^\prime_2\rho\pi}$ & $0.106$ & $0.102$\\ \hline

$\chi^2_{\pi^+\pi^-}$ ($307$ n.d.f.) & $49.9$ & $24.5$
\\ \hline

$\chi^2_{\omega\pi^0}$ ($53$ n.d.f.) & $29.8$ & $28.0$
\\ \hline

$\chi^2_{\eta\pi^+\pi^-}$ ($39$ n.d.f.) & $39.6$ & $39.8$
\\ \hline

$\chi^2_{K^+K^-}$ ($78$ n.d.f.) & $62.9$ & $52.5$
\\ \hline

$\chi^2_{\pi^+\pi^-\pi^0}$ ($142$ n.d.f.) & $121.4$ & $ 133.3$
\\ \hline

$\chi^2_{tot}$/n.d.f. &
$303.7/619$ & $278.1/619$
\\ \hline

\end{tabular}
\end{center}

\vspace{10mm}

Let's consider $\phi\to\eta\pi^0\gamma$ and $\phi\to\pi^0\pi^0\gamma$ decays. In Ref. \cite{achasov-89} the $K^+K^-$ loop model of these decays was suggested, and in Ref. \cite{achasov-2003} it was shown that these decays should basically go through $s\bar s$ pair.

\begin{figure}[h]
\begin{center}
\includegraphics[width=10cm]{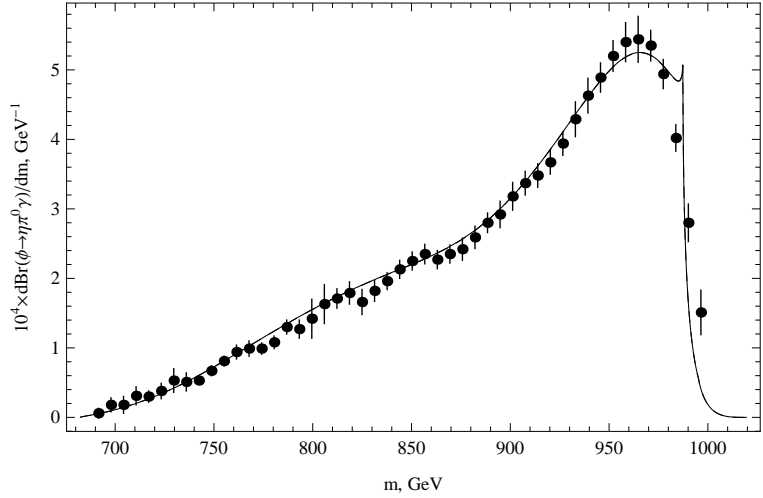}
\end{center}
\caption{The plot of mass spectra $\frac{1}{\sigma_\phi}\,\frac{d\sigma(e^+e^-\to \eta\pi^0\gamma)}{dm}$. Solid line is for Fit 1, dashed line is for Fit 2 (indistinguishable), the data on differential branching $dBr(\phi\to\eta\pi^0\gamma)/dm$ is taken from Ref. \cite{kloea02009}.}
\label{dBra0}
\end{figure}

From a more wide view, we should consider cross sections of the reactions $e^+e^-\to K^+K^-\to(a_0+a_0')\gamma\to\eta\pi^0\gamma$ and $e^+e^-\to K^+K^-\to(\sigma+f_0)\gamma\to \pi^0\pi^0\gamma$, which are proportional to $\sigma(e^+e^-\to K^+K^-)$. For example, $$\frac{d\sigma(e^+e^-\to K^+K^-\to(\sigma+f_0)\gamma\to \pi^0\pi^0\gamma,s,m)}{dm}=$$ \begin{equation}
\frac{6\pi s}{ q_{K^+K^-}^3}\,\sigma(e^+e^-\to K^+K^-,s)\,\frac{1}{g_{\phi K^+K^-}^2} \frac{d\Gamma (\phi\to K^+K^-\to\pi^0\pi^0\gamma,m)}{dm}
\end{equation}
\noindent where $m$ is the invariant mass of the $\pi^0\pi^0$ system, $d\Gamma (\phi\to K^+K^-\to\pi^0\pi^0\gamma)/dm$ was designated as $d\Gamma_S/dm$ in our previous works. The similar expression is for $e^+e^-\to K^+K^-\to(a_0+a_0')\gamma\to\eta\pi^0\gamma$ differential cross section.

On the PDG $\phi$ mass $\sqrt{s}=1019.461$ MeV \cite{pdg-2020} for Fit 2 we have $\sigma(e^+e^-\to K^+K^-, 1019.461\mbox{ MeV})=2366$ nb, and $2368$ nb for Fit 1, i.e. difference is less than 0.1 percent, practically preserving $e^+e^-\to\eta\pi^0\gamma$ and $e^+e^-\to\pi^0\pi^0\gamma$ differential cross sections. For an illustration we draw Fig. (\ref{dBra0}), where $\frac{1}{\sigma_\phi}\,\frac{d\sigma(e^+e^-\to \eta\pi^0\gamma,s,m)}{dm}$ is shown for Fits 1, 2 together with precise KLOE data on the $\phi\to\eta\pi^0\gamma$ decay (the background from the intermediate $\rho\pi^0$ state is also taken into account). One can see that the difference between Fit 1 and Fit 2 curves is indistinguishable. Here $m$ is the invariant mass of the $\eta\pi^0$ system, $\sigma_\phi=12\pi Br(\phi\to e^+e^-)/m_\phi^2=4.2$ $\mu$b \cite{pdg-2020} is the conventional Born total cross section of the $\phi$ production. The $a_0$ parameters are not far from those in Ref. \cite{ourD}, and all the data described in that work is even better described now. Since the analysis of radiative $\phi$ decays is not a topic of the given work, we don't show all formulas and don't discuss essential details concerning this illustration.

On $\sqrt{s}=1019.461$ MeV the absolute value of $\phi,\phi^{\prime}_1,\phi^{\prime}_2$ sector contribution to $e^+e^-\to K^+K^-$ amplitude is 99.2\% for Fit 2. But the $\phi$ contribution is only 16\% for Fit 2, while for Fit 1 one has 99.4\%. If we treat $\phi^{\prime}_{1,2}$ as $s\bar s$ states, the picture of Ref. \cite{achasov-2003} is kept. The real part of the $G_{\omega\varphi}(s)$ determinant vanishes at $1019.75$ Mb for Fit 2, this position may be treated as an estimation of the physical $\phi$ mass.

Some "thinking inertia" can take place because of $\phi(1020)$ narrowness. It happens because of the $K\bar K$ threshold vicinity, but it should not obligatory suppress real parts of $\Pi_{\varphi\varphi^{\prime}_{1}}$ and $\Pi_{\varphi\varphi^{\prime}_{1}}$ changing the mass. Moreover, these real parts are quite moderate in Fit 2, see Table I.

\section{Conclusion}
\label{Conclusion}

In this work a method of model accuracy estimation is proposed. The data is simultaneously described in the frame of GVDM with overall model accuracy $\beta=0.06$. The isovector channels $e^+e^-\to\pi^+\pi^-$ and $e^+e^-\to\omega\pi^0$ can be described with $\beta=0$ simultaneously with the same $\beta=0.06$ for other reactions. Note the rest isovector channel $e^+e^-\to\eta\pi^+\pi^-$ has approximations inside GVDM in our model, see remarks between Eq. (\ref{eq5}) and Eq. (\ref{yDef}).

Another way to estimate the model precision is provided by the calculation without model error subtraction, as in usual $\chi^2$ criterion. In this case we set $\beta=0$ in Eqs. (\ref{midValFull}) so $\langle a_i^2/\sigma_i^2\rangle=1+\frac{(x^0_i-x^{th}_i)^2}{2\sigma_i^2}\geq 1$. If we still take $x^0_i=x^{th}_i/(1-\beta)$ then $\langle a_i^2/\sigma_i^2\rangle=1+\Big(\frac{\beta x^{th}}{\sqrt{2}\sigma}\Big)^2$. This alternative also suffers from possible "masking effect" discussed in Sec. \ref{SecChi2}, but it allows to test the effect of non-Gaussianity of real $x^{exp}$ distribution. For Fit 1 this variant of our approach gives $\bar \chi^2_{\pi\pi}=96$, $\bar \chi^2_{\omega\pi}=36$, $\bar \chi^2_{\eta\pi\pi}=36$, $\bar \chi^2_{KK}=60$, and $\bar \chi^2_{3\pi}=132$, total $\bar \chi^2=360$. It means that the method is stable and effects of non-Gaussianity are not crucial.

In sum, on the current stage the data agree with GVDM accuracy 6\% or better. In future we plan to take into account more reactions and improve the model, including an attempt to use propagators from Ref. \cite{AchKozh-2013}.

An interesting effect considered in this work is the difference of bare and physical mass of the narrow $\phi$ meson. While optimal bare mass is about 1028 MeV, after taking mixing into account we have physical mass near the Ref. \cite{pdg-2020} value.

\section{Acknowledgement}

The study was carried out within the framework of the state contract of the Sobolev Inst. of Math. (project no. 0314-2019-0021).

\end{document}